# Drastic enhancement of the Raman intensity in few-layer InSe by uniaxial strain

Chaoyu Song[1,3†], Fengren Fan[1,2,3†], Ningning Xuan[4†], Shenyang Huang[1,3], Chong Wang[1,3], Guowei Zhang[1,3], Fanjie Wang[1,3], Qiaoxia Xing[1,3], Yuchen Lei[1,3], Zhengzong Sun[4*], Hua Wu[1,2,3*], Hugen Yan[1,3*]

1. State Key Laboratory of Applied Surface Physics and Department of Physics, Fudan University, Shanghai 200433, China.

2. Key Laboratory of Computational Physical Sciences (Ministry of Education), Fudan University, Shanghai 200433, China

3. Collaborative Innovation Center of Advanced Microstructures, Nanjing 210093, China

4. Department of Chemistry and Shanghai Key Laboratory of Molecular Catalysis and Innovative Materials, Fudan University, Shanghai 200433, China

† These authors contributed equally

* Emails: hgyan@fudan.edu.cn (H. Y.), wuh@fudan.edu.cn (H. W), zhengzong_sun @fudan.edu.cn (Z. S.)

**ABSTRACT**

The vibrational and electronic properties of 2-dimensinal (2D) materials can be efficiently tuned by external strain due to their good stretchability. Resonant Raman spectroscopy is a versatile tool to study the physics of phonons, electrons and their interactions simultaneously, which is particularly useful for the investigation of strain effect on 2D materials. Here, for the first time, we report the resonant Raman study of strained few-layer InSe (γ-phase). Under ~ 1% of uniaxial tensile strain, one order of magnitude enhancement of Raman intensity for longitudinal optical (LO) phonon is observed, while other modes exhibit only modest change. Further analysis demonstrates that it arises from the intraband electron-phonon scattering channel for LO phonon in resonance. The large enhancement of Raman intensity provides us a sensitive and novel method to characterize the strain effect and a mapping of the strain distribution in a wrinkled sample is demonstrated. In addition, we observed sizable redshifts of first-order optical phonon modes. The shift rate exhibits phonon mode dependence, in excellent agreement with density functional perturbation theory (DFPT) calculations. Our study paves the way for sensitive strain quantification in few-layer InSe and its application in flexible electronic and optoelectronic devices.

# I. INTRODUCTION

Mechanical cleavage of graphene [1] by K. S. Novoselov et al. arouses tremendous research interest in 2D materials. A variety of 2D semimetals and semiconductors have been discovered ever since, such as transition metal dichalcogenides (TMDCs) [2], silicone [3], stanine [4] and black phosphorus [5,6]. Atomically thin indium selenide (γ-phase) joins the family lately with unique electronic properties [7-9]. Quantum Hall effect was observed in the high quality few-layer InSe electronic devices [9]. Strong quantum confinement in the out-of-plane direction gives rise to layer-dependent bandgap [7], covering a large range of visible and near infrared regions. Few-layer InSe has promised great application potentials in electronics and optoelectronics [10-12].

The mechanical stretchability of 2D materials opens the door for straining, to continuously and reversibly tune their lattice constants and electronic properties [13]. Raman spectroscopy is a crucial diagnostic tool to evaluate the strain effect. Phonon softening and splitting are commonly observed in 2D materials under uniaxial tensile strain, such as graphene [14,15], TMDCs [15,16] and black phosphorus [18,19], indicating the weakening of the bond strength and the symmetry-breaking. The band structure and electronic properties of 2D materials can be engineered efficiently via strain as well. For example, prominent strain-induced shift of the band gap and indirect-to-direct bandgap transition were observed in multilayer TMDCs [20]. Owing to the small Young's modulus ( ~ 45 N/m) [21], the bandgap of few-layer InSe can be easily tuned by uniaxial tensile strain with shift rate up to 90-150meV/% [22,23]. Therefore, few-layer InSe, with largely strain-tunable bandgaps, exhibits great application potentials in flexible and wearable electronics. It calls for comprehensive understanding of lattice vibrations and their interactions with electrons in strained InSe.

In this work, for the first time, we performed resonant Raman spectroscopy to study the vibrational and electronic properties of few-layer InSe (γ-phase) under uniaxial tensile strain. The Raman intensity is particularly focused on, in contrast to previous similar studies on other 2D materials, in which the Raman intensity typically does not change notably under strain and little attention has been paid to. The energy match between photons and electronic intermediate states (exciton or continuum bands) provides us abundant information on the band structures and electronic properties, in addition to the phonon physics [24]. Combing the merits of tunable strain and resonant Raman spectroscopy, we are able to measure the Raman spectra of few-layer InSe under resonant and near-resonant conditions with a single 514.5 nm laser. Drastic enhancement of the Raman intensity of $A_1$ ($\Gamma_1^1$)-LO and $E$ ($\Gamma_3^1$)-LO phonon mode is observed in the process of applying uniaxial tensile strain from zero to 1.15%, while that for other Raman modes only changes modestly. This is ascribed to the intraband electron-phonon scattering channel for LO phonon in resonance. In addition, sizable redshifts are observed for first-order phonon modes located in the range of 100-300 $cm^{-1}$. Furthermore, as a proof of concept, we determine the spatial variation of strain in a wrinkled few-layer InSe flake by mapping the Raman intensity of $A_1$ ($\Gamma_1^1$)-LO phonon.

## II. EXPERIMENTAL METHODS

Few-layer InSe was exfoliated from bulk crystal of InSe (2D Semiconductors Inc.) onto polydimethylsiloxane (PDMS) substrates by scotch tape method [1]. The flakes were subsequently transferred to flexible PETG substrates with thickness of 0.3 mm. The thickness of InSe was determined by optical contrast , and further verified by Atomic force microscope (AFM) and Photoluminescence (PL) spectroscopy (see the Supplemental Material (SM) sections S1 [25]) The wrinkled InSe was fabricated by conventional method of elongating and releasing flexible substrate [26,27]. The PDMS substrate (1mm thick) was firstly prestretched by 50-100%. Then InSe flakes were deposited. The PDMS substrate was then released suddenly, creating wrinkles in few-layer InSe (see SM sec. S10 [25]). The topography of the wrinkles was measured by Bruker Dimension Edge AFM (probe model RTESP-300, tapping mode).

Resonant Raman spectra were taken by Horriba HR800 Raman system with the excitation of 514.5 nm laser (spot size ∼1μm, laser power of 50 μW). The incident laser was focused by a 100× (N.A.=0.9) objective. The PL measurement was performed by the same setup with a 473 nm laser (spot size ∼1μm, laser power of 500 μW) for excitation. The laser power was kept low enough to prevent potential damage to the samples. The differential reflectance spectra were taken by an Andor SR500i spectrometer in conjuction with a Nikon Eclipse Ti microscope equipped with a 50× (N.A.=0.7) objective. All experiments reported in this paper were conducted at room temperature.

## III. RESULTS AND DISCUSSIONS

### A. The vibrational properties of unstrained few-layer InSe

InSe (γ-phase) crystal has a layered hexagonal structure as depicted in Fig. 1(a). In each layer, the lattice consists of four atomic planes arranged in the sequence of Se-In-In-Se. Individual layers are held together by van der Waals forces. Few-layer InSe (γ-phase) belongs to point group $C_{3v}$, and the normal modes of vibrations are $4A_1$ ($\Gamma_1$)+$4E$ ($\Gamma_3$). Among them, one $A_1$ mode and one E mode are acoustic while the remaining modes are optical (all both infrared and Raman active) [28]. The vibrations are illustrated in Fig. 1(b). Fig. 1(c) shows Raman spectra of unstrained 3-, 10-, 30-layer and bulk InSe with the excitation of 514.5 nm laser. The E ($\Gamma_3^2$) mode is beyond the lower limit of our setup, hence not discussed here. Five Raman peaks are observed in the range of 100-300 cm$^{-1}$. They are $A_1$ ($\Gamma_1^2$) at 115 cm$^{-1}$, E ($\Gamma_3^3$) / E ($\Gamma_3^1$) -TO at 178 cm$^{-1}$ (TO is for transverse optical phonon), $A_1$ ($\Gamma_1^1$) -TO at 196 cm$^{-1}$, $A_1$ ($\Gamma_1^1$) -LO at 199 cm$^{-1}$ and $A_1$ ($\Gamma_1^3$) at 227 cm$^{-1}$, respectively. All Raman spectra reported in this work are measured by back scattering geometry. The $A_1$ ($\Gamma_1^1$)-LO and E ($\Gamma_3^1$)-LO modes are forbidden or extremely weak by selection rules in such geometry [29-31], hence not observed in off-resonant conditions with the excitation of 532 nm (see SM sec. S5 [25]) and 633 nm lasers [11]. However, the intensity of Raman scattering, especially for the forbidden modes, is enhanced if the energy of incident photon (514.5 nm，2.41 eV) is close enough to that of B-transition, which is the electronic transition between the top of Se-4px/y orbital band and the bottom of conduction band (In-5s orbital) [32]. Therefore, we could observe a weak peak originated from $A_1$ ($\Gamma_1^1$)-LO phonon mode as shown in Fig. 1(c).

**B. Giant enhancement of the Raman intensities of LO phonons**

Uniaxial tensile strain was applied to few-layer InSe by two-point bending scheme as shown in Fig. 2(a). Flexible poly (ethylene terephthalate)-glycol (PETG) substrate was chosen, such that minimal background signal was from the substrate itself. The strain in the stretch direction is determined according to the deflection of PETG substrates [20] (see SM sec. S2 [25]), and the strain along the other direction is ignored in the analysis. The comparison of strain effect without and with clamping shows that there is no sliding in InSe due to the good stretchability of the sample (see SM sec. S12 [25]). Therefore, clamping is not necessary for our study with moderate strain values.

We systematically measured the evolution of the resonant Raman spectra of few-layer InSe under uniaxial tensile strain. Two major strain effects are observed. One is the large enhancement of $A_1$ ($\Gamma_1^1$)-LO and E ($\Gamma_3^1$)-LO phonon modes, and the other is the sizable redshift of all first-order Raman modes. We categorize our samples into two groups with thickness of 3-6 layers and 10-15 layers according to the different enhancement magnitude, as shown in Fig. 2(b) and 2(c). It is worth noting, however, that all first-order phonon modes should exhibit enhancement in resonance conditions [29], but the enhancement of the $A_1$ ($\Gamma_1^1$)-LO and E ($\Gamma_3^1$)-LO modes is one order of magnitude larger than that of other modes. Therefore, in this work, we mainly focus on the abnormally large enhancement of those two modes, and all spectra in Fig. 2(b) and 2(c) are normalized to have the same intensity for $A_1$ ($\Gamma_1^2$) mode. As shown in Fig. 2(b) and 2(c), the intensity of $A_1$ ($\Gamma_1^1$)-LO mode increases dramatically when uniaxial tensile strain is applied from zero to 1.15%. For 10-15 layers InSe, a broad shoulder simultaneously emerges at the high energy side of the $A_1$ ($\Gamma_1^1$)-LO Raman peak, which is assigned to E ($\Gamma_3^1$)-LO mode [28].

By fitting the spectra with Lorenz formula, we obtain the intensity, frequency, and width of each Raman peak. Fig. 2(d) and 2(e) show the evolution of normalized intensity of the enhanced phonon modes as uniaxial tensile strain is applied. The data points and error bars are based on the measurements of more than 10 samples for each group. For comparison, the normalized intensity of $A_1$ ($\Gamma_1^3$) mode is also plotted. The normalized intensity of $A_1$ ($\Gamma_1^3$) mode almost remains a constant, showing that the ratio of the Raman intensity of $A_1$ ($\Gamma_1^3$) to $A_1$ ($\Gamma_1^2$) is the same in the process of applying strain. Therefore these selection-rule-allowed phonon modes exhibit a similar modest enhancement factor, which behave quite differently from the forbidden LO phonon modes. We use the enhancement ratio $\alpha = \dfrac{I\left(with\ 1.15\%\ tensile\ strain\right)}{I\left(without\ strain\right)}$, which is the ratio of the normalized Raman intensity under 1.15% tensile strain to that without strain, to quantify the enhancement magnitude. For 3-6 -layer InSe, the enhancement ratio α of $A_1$ ($\Gamma_1^1$)-LO is 5±1. For 10-15 layers InSe, the enhancement effect becomes even more pronounced, with the enhancement ratio α of 18±4. The Raman peak of E ($\Gamma_3^1$)-LO mode is absent without strain, and gradually emerges with strain >0.6%. By assuming that the intensity of E ($\Gamma_3^1$)-LO mode without strain is at the same level of the spectrum background, a conservative estimation gives the enhancement ratio α of over 10.

**C. Enhancing Raman intensities by intraband electron-phonon scattering channel**

It is well-known that the intensity of Raman scattering is enhanced under resonance conditions [24]. For few-layer InSe, the strain-induced modification of band structure [22, 23] evokes the change of resonant conditions when the excitation laser wavelength is fixed. As shown in Fig. 3(a), when uniaxial tensile strain is applied, the energy of B-transition shifts closer to that of incident photon (514.5 nm, 2.41 eV) and eventually the intermediate electron/hole state changes from virtual to real energy state, so the Raman intensity exhibits enhancement. PL measurements were performed to quantitatively illustrate the strain effect on the band structures. The position of PL peaks corresponds to the energy of exciton, which is slightly smaller than the energy of quasi-particle electronic transition with the difference as the exciton binding energy [33]. The electronic intermediate state of resonant Raman scattering of few-layer InSe is the exciton state associated with the B-transition (we call it exciton B for simplicity). Fig. 3(b) shows the PL spectra of a strained 12-layer InSe, the peak shifts from 2.46 eV to approximately 2.41 eV as uniaxial tensile strain is applied from zero to 1.15%. Therefore, the resonance conditions are gradually satisfied. For few-layer InSe, the strain effect on B-transition is almost the same, all exhibiting similar shift rate of (45±10) meV/% according to the PL spectra of strained 8-layer and 30-layer (see SM sec. S6 [25]). From the above argument, it is apparent that such enhancement of Raman intensity should be laser-wavelength dependent. If the laser energy is above the B-transition, tensile strain will separate them further and no enhancement is supposed to occur. Indeed, this is the case for the excitation with a 473 nm laser, as shown in SM sec. S4 [25]. In addition, we used 488nm laser (2.54eV) to excite Raman scattering of a 4-layer InSe, whose B-transition energy is approximately 2.53eV, and resonance effect was observed as illustrated in SM sec. S11 [25].

The enhancement shows certain layer-dependence as well, since the B-transition energy depends on the sample thickness. The different enhancement ratio between 3-6 layers and 10-15 layers InSe is consistent with this scenario. Fig. 3(c) shows the differential reflectance spectra $\frac{\Delta R}{R} = \frac{R_{sample} - R_{substrate}}{R_{substrate}}$ of 1-layer to 8-layer InSe, the bump position (indicated by triangles) of the differential reflectance corresponds to exciton B [2]. Strong interlayer interactions give rise to prominent layer-dependent energy of exciton B, which is 2.47eV for 8-layer and increases to about 2.81eV for monolayer (without strain). Combing with the results of PL measurements, we summarize the layer- and strain-dependent energy of the exciton B in Fig. 3(d). It qualitatively explains why the 10-15 layers InSe exhibits larger resonant effect than the 3-6 -layer InSe. The energy of exciton B is around 2.45-2.47 eV for 10-15 layers without strain, and 1.15% uniaxial tensile strain makes it redshift by 50 meV, in perfect match with the incident photon energy. While for unstrained 3-6 layers InSe, the energy of exciton B is about 2.50-2.55 eV. Therefore, tensile strain will make the resonant effect enhance but still away from the optimum even for maximal strain (1.15%, redshift of ~ 50 meV).

The enhancement depends on the phonon mode character. On the contrary to the conventional resonant Raman scattering that all phonon modes are enhanced on the same footing, herein we only observe a drastic enhancement of $A_1$ ($\Gamma_1^1$)-LO and $E$ ($\Gamma_3^1$)-LO modes, while the enhancement of other phonon modes is much less significant by contrast. The larger enhancement of LO modes arises from the dominant intraband Fröhlich electron-phonon interactions near the resonance, which is originated from the scattering of electron or hole within the same band by the macroscopic electric field of an LO phonon [24]. According to Fermi's golden rule, the cross section of Raman scattering reads

$$\sigma \cong |W_{fi}|^2 = \left| \sum_\alpha \frac{M_{eR}(\omega_s) M_{ep} M_{eR}(\omega_i)}{(\hbar\omega_s - \hbar\omega_\alpha + i\Gamma)(\hbar\omega_i - \hbar\omega_\alpha + i\Gamma)} \right|^2 \quad (1)$$

where $W_{fi}$ is the matrix elements for Raman scattering, $\omega_i$, $\omega_s$ and $\omega_\alpha$ are the frequencies of incident photon, scattered photon and electronic intermediate state (the frequency of first-order phonon is $\omega_0 = \omega_i - \omega_s$ for Stokes Raman scattering) respectively, $\Gamma$ is the damping factor of exciton, $M_{eR}$ and $M_{ep}$ denote the matrix elements for the electron-photon and electron-phonon interactions, respectively.

For selection-rule-allowed phonon modes, the phonon matrix elements $M_{ep}$ is non-vanishing and independent of phonon wave vector $q$ as $q \to 0$. The two denominators account for the enhancement of Raman cross section in resonance conditions [34]. On the contrary, for the forbidden Raman modes, the phonon matrix element $M_{ep}$ for intraband Fröhlich scattering vanishes at long wavelength ($q \to 0$), so the higher order term of the wave vector dominates the scattering cross section. The $q$-dependent matrix element $M_{ep}$ is proportional to $qa^2$, where $a$ is the energy-dependent characteristic length [24,35]: $a \propto \dfrac{1}{\left|\hbar\omega_s - \hbar\omega_\alpha + i\Gamma\right|^{1/2} + \left|\hbar\omega_i - \hbar\omega_\alpha + i\Gamma\right|^{1/2}}$. Therefore, in addition to the enhancement caused by the two denominators in equation (1), the forbidden LO phonons exhibit more pronounced enhancement than the allowed modes by intraband electron-phonon scattering channel. Such effect was also observed in CdS [35,36], InSb [37] and bulk γ-InSe [29,30] previously.

**D. Phonon softening of strained InSe**

Phonon softening under tensile strain is a common phenomenon in 2D materials [14-19]. As tensile strain is applied, the length of the covalent bonds increases, consequently the restoring force in the vibrations becomes weaker and the phonon frequency decreases. As shown in Fig. 2(b) and 2(c), except for a tiny shift of $A_1$ ($\Gamma_1^2$) mode, sizable redshifts of all other phonon modes are observed. The phonon frequency and the corresponding redshift for each phonon mode exhibit no meaningful layer dependence, therefore for simplicity, in Fig. 4(a)-(e), we only show the evolution of Raman peak positions of $A_1$ ($\Gamma_1^1$), E ($\Gamma_3^3$) / E ($\Gamma_3^1$)-TO, $A_1$ ($\Gamma_1^1$)-LO, E ($\Gamma_3^1$)-LO and $A_1$ ($\Gamma_1^3$) modes of strained 10-15 layers InSe. Similar data for strained 3-6 layers samples are presented in SM sec. S3 [25]. The shift rates under strain are summarized in Table 1. The shift rates of $A_1$ ($\Gamma_1^1$)-LO, E ($\Gamma_3^1$)-LO and $A_1$ ($\Gamma_1^3$) modes are much larger than that of $A_1$ ($\Gamma_1^2$) mode, because the $A_1$ ($\Gamma_1^2$) mode is the breathing mode of two In-Se sublayers and in-plane strain has weaker influence on the atomic vibration in the out-of-plane direction (see SM sec. S8 [25]). We performed first-principle calculation [38-41] to obtain the phonon dispersion of InSe. The uniaxial strain is introduced by expanding the lattice along the zigzag direction (a-axis) up to 2% with 0.5% per step, while the same strain along the other (armchair) direction gives similar results. SM sec. S7 [25] shows phonon dispersion of monolayer InSe with and without strain, which confirms that all first order optical phonon modes exhibit sizable redshifts as uniaxial tensile strain is applied. The calculated shift rates are summarized in Table 1, which are consistent with experimental results.

### E. Strain distribution through Raman mapping

The strong dependence of Raman intensity on strain can be utilized as a sensitive means to determine the spatial variation of strain. As a proof of concept, we mapped the $A_1$ ($\Gamma_1^1$)-LO phonon mode intensity for a wrinkled sample. We intentionally formed wrinkles in few-layer InSe and induced local uniaxial tensile strain [26,27]. As shown in Fig. 5(a) and 5(c), two wrinkles were created on a 15-layer InSe (see SM sec. S10 [25]). AFM topography (see Fig. 5(d)) shows that the height of wrinkles is about 2-3 μm and the width is about 1 μm. According to Fig. 5(b) and 5(g), the normalized Raman intensity of $A_1$ ($\Gamma_1^1$)-LO phonon mode on the wrinkle is one order of magnitude larger than that in flat region, indicating stronger resonance effect on top of the wrinkle. The spatial distribution of normalized Raman intensity is shown in Fig. 5(e). In addition, the strain effect can also be mapped by phonon frequency as illustrated in Fig. 5(f) and 5(g). The Raman peak of $A_1$ ($\Gamma_1^1$)-LO phonon mode redshifts by approximately 0.8-1.2

cm$^{-1}$, and a maximum of 0.3-0.5% strain was induced based on Table 1 (shift rate -2.5±0.1 cm$^{-1}$/%). The spatial variation of Raman intensity provides us a novel method to map the strain of 2D materials. Compared to strain-induced shift of Raman peaks, large intensity contrast can potentially contribute to better mapping sensitivity and contrast for the strain field, if the photon energy is close to the exciton energy. Through the comparison of Fig. 5(d)-(f), we find that the profile of wrinkles is perfectly mapped by the spatial distribution of Raman intensity, while that obtained from phonon frequency exhibits larger blurring. This is because the Raman intensities of A$_1$ ($\Gamma_1^1$)-LO phonon mode between wrinkled and flat area are distinctively difference (an order of magnitude), while the shift of peak position is small (<1 cm$^{-1}$) and demands careful differentiation.

## IV. CONCLUSION

In summary, we measured the resonant Raman spectra of few-layer InSe. Large enhancement of phonon modes A$_1$ ($\Gamma_1^1$)-LO and E ($\Gamma_3^1$)-LO is observed, which originates from the dominant intraband Fröhlich electron-phonon interactions. Phonon softening is obseved in the strained InSe and the shift rates are perfectly consistent with DFPT calculations. The spatial distribution of strain in a wrinkled few-layer InSe is illustrated by mapping the Raman intensity of forbidden A$_1$ ($\Gamma_1^1$)-LO phonon mode.

**ACKNOWLEDGMENTS**

C. S., F.F. and X.N. contributed equally to the work. H.Y. is grateful to the financial

support from the National Young 1000 Talents Plan, National Natural Science Foundation of China (grants: 11874009, 11734007), the National Key Research and Development Program of China (grants: 2016YFA0203900, 2017YFA0303504), Strategic Priority Research Program of Chinese Academy of Sciences (XDB30000000), and the Oriental Scholar Program from Shanghai Municipal Education Commission. Part of the experimental work was carried out in Fudan Nanofabrication Lab. H. W. is supported by the National Natural Science Foundation of China (grants: 11474059, 11674064), and by the National Key Research and Development Program of China (grant: 2016YFA0300700). Z. Sun is grateful to the financial support from the National Key Research and Development Program of China (grants: 2016YFA0203900, 2017YFA0207303), the National Natural Science Foundation of China (grants: 21771040) and the 1000 Plan Program for Young Talents. C.W. is grateful to the financial support from the National Natural Science Foundation of China (grants: 11704075) and China Postdoctoral Science Foundation. G.Z. acknowledges the financial support from the National Natural Science Foundation of China (grant: 11804398).

**TABLE 1.** The shift rates (in cm$^{-1}$/%) of few-layer InSe are based on Raman experiments and monolayer InSe from DFPT calculations. The point group of monolayer InSe is D$_{3h}$, and the corresponding irreducible representations were denoted.

| Mode | 3-6L | 10-15L | Mode | 1L |
|---|---|---|---|---|
| E ($\Gamma_3^2$) | | | E'' | -0.4 |
| A$_1$ ($\Gamma_1^2$) | -0.8±0.1 | -0.8±0.1 | A$_1$' | -0.7 |
| E ($\Gamma_3^3$) | -1.5±0.2 | -1.5±0.1 | E'' | -1.3 |
| E ($\Gamma_3^1$)-TO | -1.5±0.2 | -1.5±0.1 | E'-TO | -1.4 |
| A$_1$ ($\Gamma_1^1$)-LO | -2.7±0.2 | -2.5±0.1 | A$_2$''-LO | -2.5 |
| E ($\Gamma_3^1$)-LO | | -2.0±0.5 | E'-LO | -1.1 |
| A$_1$ ($\Gamma_1^3$) | -2.4±0.2 | -2.5±0.1 | A$_1$' | -1.7 |

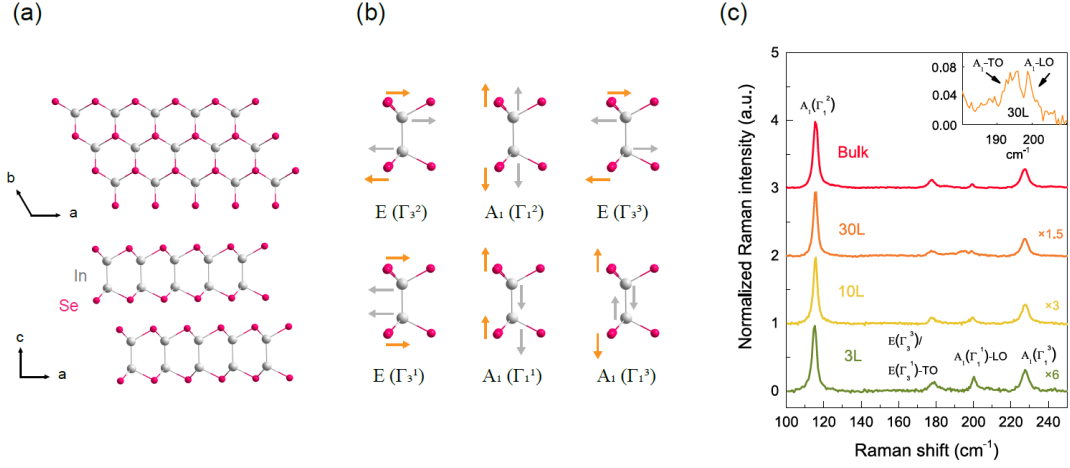

**FIG. 1.** (a) Crystal structures of monolayer InSe viewed from c-axis direction (left) and of bilayer InSe viewed from b-axis direction (right). (b) Lattice vibrations of all first-order optical phonon modes, Mulliken symbols are used in this work, while the corresponding Γ symbols are labelled in parenthesis for comparison. (c) Raman spectra of 3-layer, 10-layer, 30-layer and bulk InSe under no strain with the excitation of a 514.5 nm laser, the inset shows the zoom-in of the spectra near 200 cm$^{-1}$ for the 30-layer InSe. The spectrum intensity is multiplied by a factor indicated beside each spectrum for 3L, 10L and 30L.

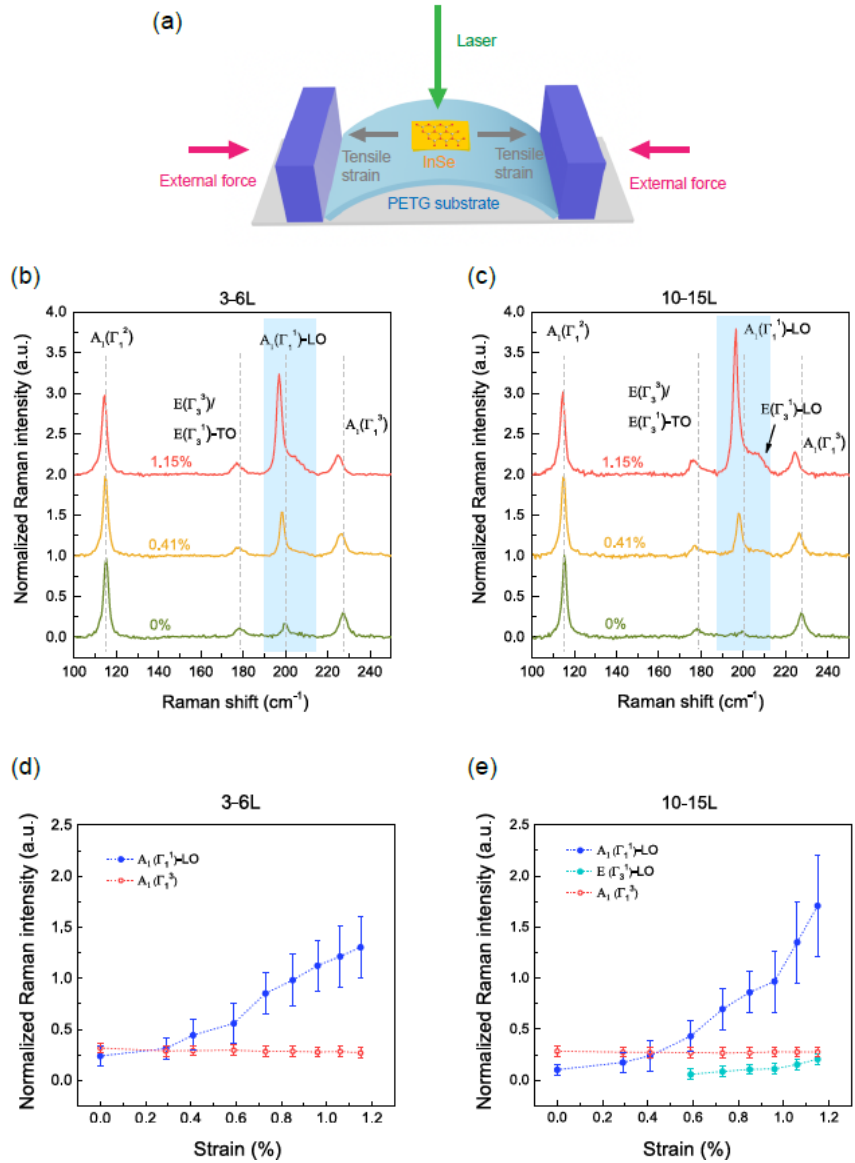

**FIG. 2.** (a) Schematic illustration of the two-point bending apparatus. Raman spectra of (b) 3-6 layers and (c) 10-15 layers InSe as uniaxial tensile strain is applied from zero to 1.15%. The assignment of corresponding phonon mode for each Raman peak is marked. The blue rectangles highlight the enhanced LO phonon mode, and the vertical grey dashed lines indicate the positions of respective Raman peaks under no strain. (d) and (e) Normalized Raman intensity of enhanced LO modes and selection-rule-allowed $A_1$ ($\Gamma_1^3$) mode for 3-6 layers and 10-15 layers InSe respectively, as a function of strain.

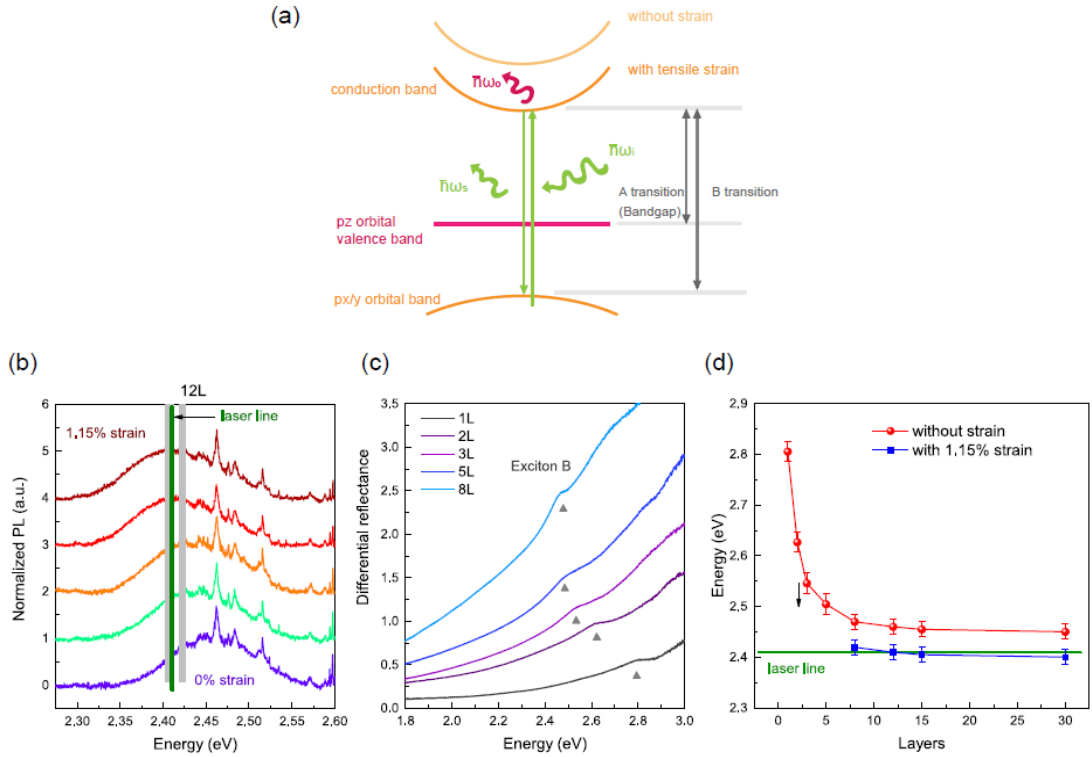

**FIG. 3.** (a) Schematic illustration of band structures and resonant Raman scattering process of InSe. The band gap transition is marked as A-transition, while the electronic transition between the top of Se-4$p_{x/y}$ orbital band and the bottom of conduction band (In-5s orbital) is marked as B-transition. (b) Normalized PL spectra of a strained 12-layer InSe, the green line is the laser line of 514.5 nm laser (2.41 eV) for Raman measurements, the two grey rectangles correspond to the sharp Raman peaks from the substrate, which we intentionally removed. Spectra are vertically shifted for clarity. (c) The differential reflectance spectra of 1- to 8- layer InSe, the grey triangles denote the absorption peak of exciton B. (d) The layer-dependent energy of exciton B without and with 1.15% uniaxial tensile strain, the data points for 8- to 30- layer InSe come from PL measurements, while the results of 1- to 5- layer InSe are originated from the differential reflectance spectra, the green horizontal line is the photon energy of the

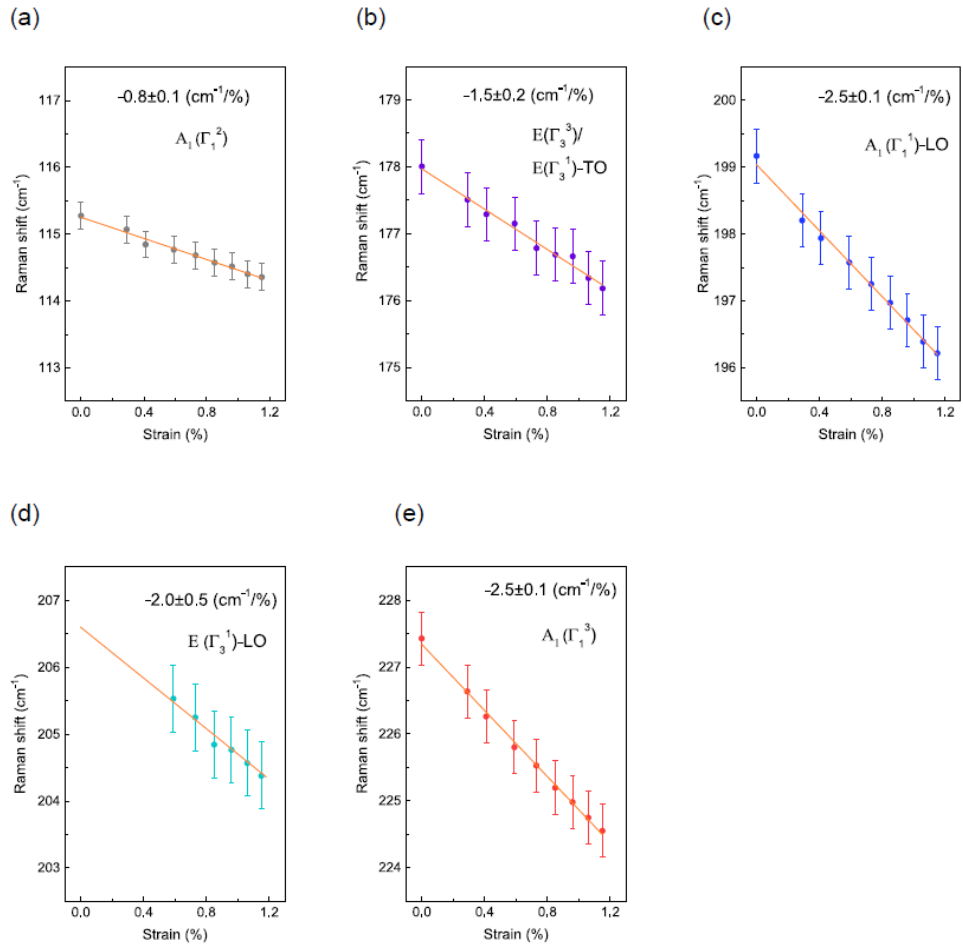

**FIG. 4.** Evolution of phonon frequencies for 10-15 layers InSe under uniaxial tensile strain. (a) - (e) correspond to the $A_1$ ($\Gamma_1^2$), E ($\Gamma_3^3$) / E ($\Gamma_3^1$) -TO, $A_1$ ($\Gamma_1^1$)-LO, E ($\Gamma_3^1$)-LO and $A_1$ ($\Gamma_1^3$) modes, respectively, in the order of increasing phonon frequency.

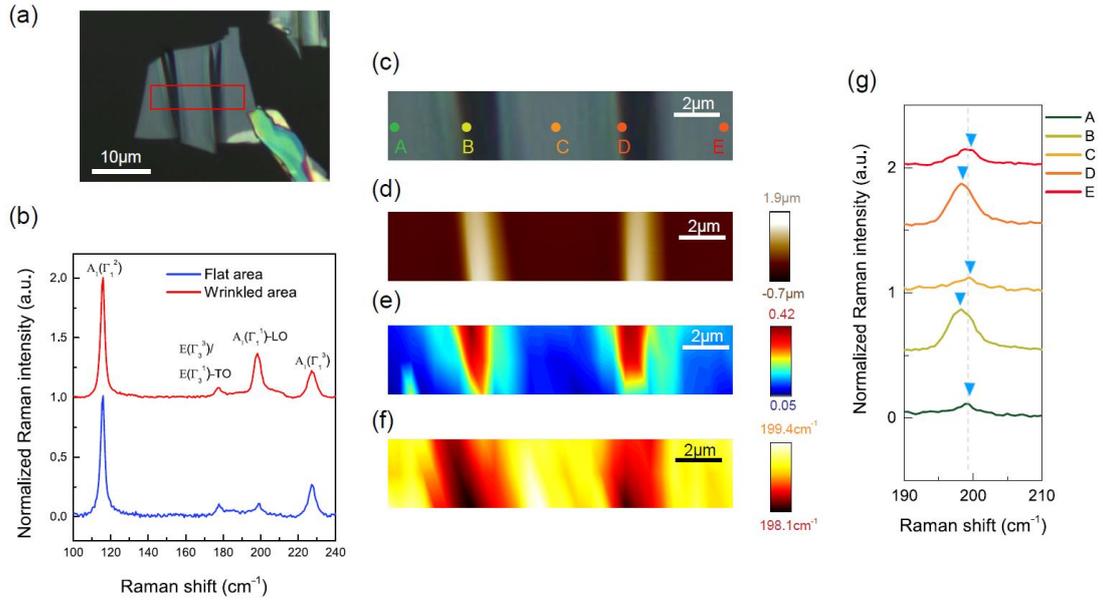

**FIG. 5**. (a) The microscope optical image of a wrinkled InSe flake (about 15 Layers), the red rectangle denotes the mapping area. (b) The Raman spectra measured on the flat and wrinkled area. (c) The optical image of wrinkled InSe with two wrinkles, the colored points A to E correspond to different flat and wrinkled areas. (d) The AFM topography image of the sample. (e) The mapping of normalized Raman intensity and (f) peak position of $A_1$ ($\Gamma_1^1$)-LO phonon mode. (g) The Raman spectra of $A_1$ ($\Gamma_1^1$)-LO phonon mode for the five points indicated in (c), the blue triangles denote the corresponding Raman peaks, the vertical grey dashed line indicates the peak position of $A_1$ ($\Gamma_1^1$)-LO phonon mode on flat area (199.0 cm$^{-1}$).

Supplemental Material

# Drastic enhancement of the Raman intensity in few-layer InSe by uniaxial strain

Chaoyu Song[1, 3†], Fengren Fan[1, 2, 3†], Ningning Xuan[4†], Shenyang Huang[1, 3], Chong Wang[1, 3], Guowei Zhang[1, 3], Fanjie Wang[1, 3], Qiaoxia Xing[1, 3], Yuchen Lei[1, 3], Zhengzong Sun[4*], Hua Wu[1, 2, 3*], Hugen Yan[1, 3*]

5. State Key Laboratory of Applied Surface Physics and Department of Physics, Fudan University, Shanghai 200433, China.

6. Key Laboratory of Computational Physical Sciences (Ministry of Education), Fudan University, Shanghai 200433, China

7. Collaborative Innovation Center of Advanced Microstructures, Nanjing 210093, China

8. Department of Chemistry and Shanghai Key Laboratory of Molecular Catalysis and Innovative Materials, Fudan University, Shanghai 200433, China

† These authors contributed equally

* Emails: hgyan@fudan.edu.cn (H. Y.), wuh@fudan.edu.cn (H. W), zhengzong_sun @fudan.edu.cn (Z. S.)

## S1. The optical contrast and layer number of few-layer InSe

The optical contrast of monolayer and bilayer InSe on PDMS is approximately 10% and 20% respectively, as illustrated in Fig. S1 (a-b). For trilayer InSe, it is about 30%, and increases at the step of 20% with the increase of each layer. The optical contrast measurements were performed through the green channel of a Nikon Eclipse-Ti Inverted microscope with a 50x objective lens and a CCD camera. The thickness determined from the optical contrast was further verified by AFM and PL spectra. S1 (c-f) show the optical image, optical contrast, AFM profile and PL spectra of a sample with 5- and 6- layer InSe. The PL of bandgap transition exhibit strong layer-dependence according to Denis A. Bandurin et al. [1] and our previous study [2]. The sample is finally transferred to silicon substrate with 300nm of silicon dioxide to perform AFM measurements. The height of one layer InSe is about 0.9 nm.

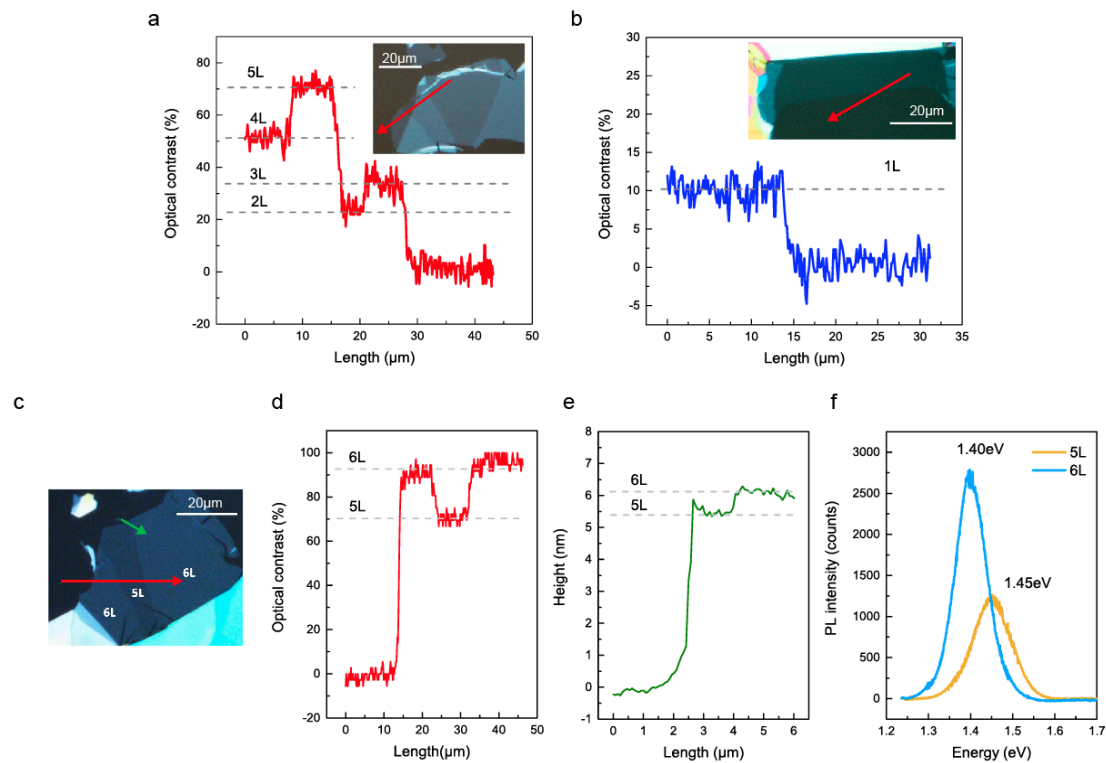

**FIG. S1.** The optical contrast of (a) multiple layer and (b) monolayer InSe, the insets

27 / 41

are the sample images and the optical contrast are shown along the red arrow ; (c) The image of a sample with 5- and 6- layer InSe, (d) the optical contrast is shown along the red arrow, (e) the AFM profile is shown along green arrow, (f) the PL spectra of bandgap transition are measured in 5- and 6- layer region separately.

**S2. The method to determine strain**

The strain ε is determined according to the deflection of the substrate [3]: $\varepsilon = \dfrac{d \sin\theta}{L}$ as illustrated in Fig. S2. Here, ε is the strain, d is the thickness of the substrate, θ is the angle between the tangent line at the end of the substrate and the horizontal line, and L is the distance between the two ends of bent PP.

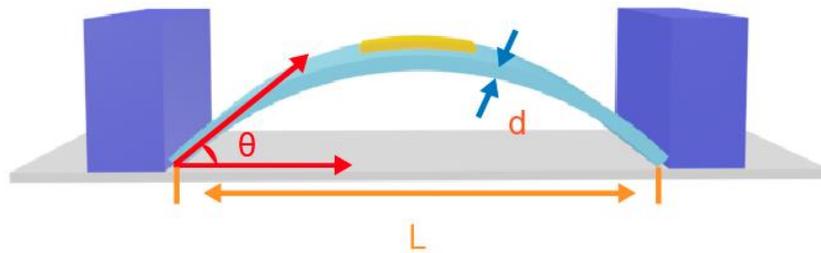

**FIG. S2.** Schematic illustration of the setup and parameters to determine the strain

## S3. The evolutions of phonon frequencies of strained 3-6 layers InSe.

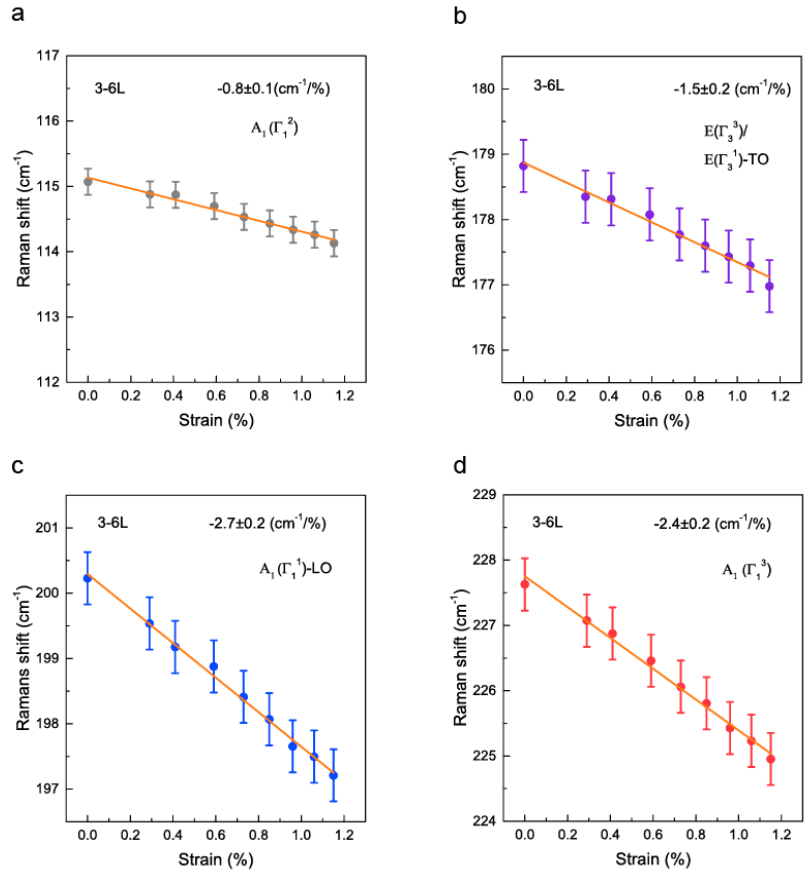

**FIG. S3.** (a)-(d) The evolution of phonon frequencies for strained 3-6 layers InSe with the excitation of a 514.5nm laser

## S4. The vibrational properties of strained few-layer InSe with the excitation of 473nm laser

We used 473nm laser to excite the Raman scattering of few-layer InSe under tensile strain. The $A_1$ ($\Gamma_1^2$) mode at about 115cm$^{-1}$ was not observed due to the limit of edge filters for 473nm laser (the lower limit is about 140 cm$^{-1}$). Therefore, here we only focus on the E ($\Gamma_3^3$) / E ($\Gamma_3^1$)-TO, $A_1$ ($\Gamma_1^1$)-LO and $A_1$ ($\Gamma_1^3$) modes, and the Raman spectra are normalized to the intensities of corresponding $A_1$ ($\Gamma_1^3$) modes. As shown in Fig. S4 (a), the intensity of $A_1$ ($\Gamma_1^1$) -LO mode does not enhance in the process of applying tensile strain based on our measurement of over 10 samples with thickness of 5L to 30L.

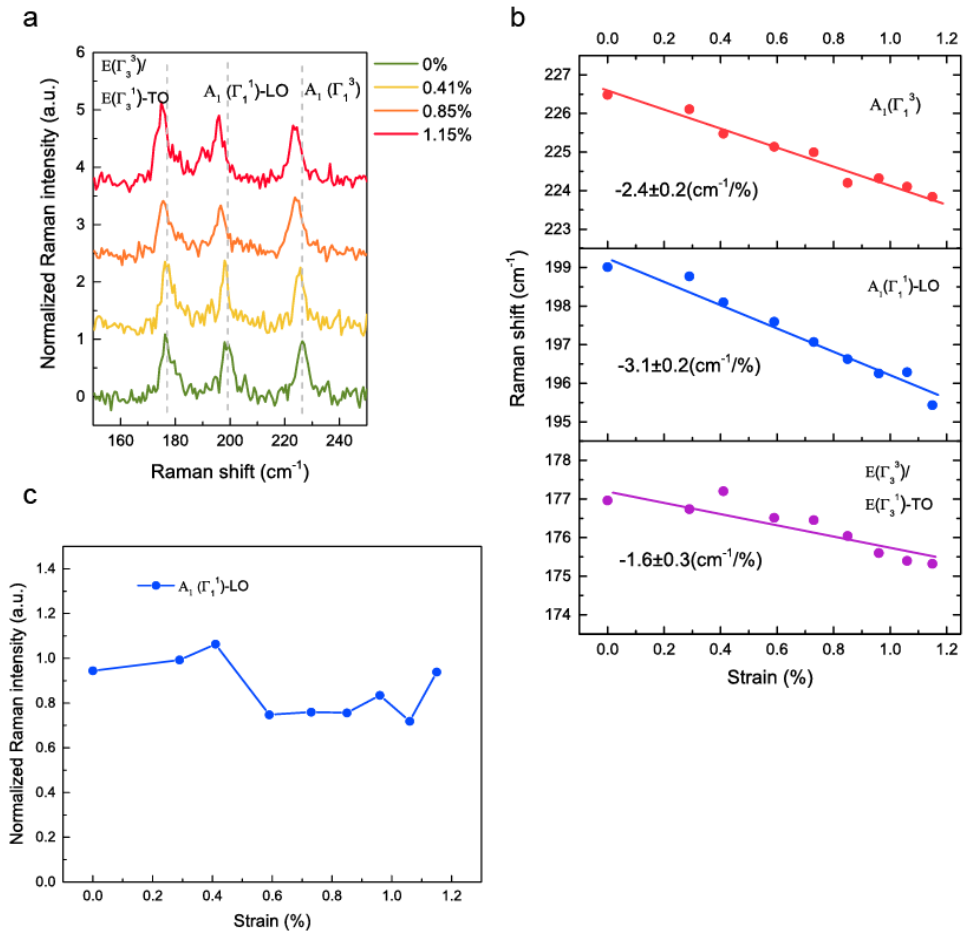

**Fig. S4.** (a) The normalized Raman spectra of a strained 8-layer InSe with the excitation

of 473nm laser. (b) The phonon frequencies and (c) the normalized Raman intensity of the $A_1$ ($\Gamma_1^1$) -LO phonon modes.

## S5. The vibrational properties of strained few-layer InSe with the excitation of 532nm laser

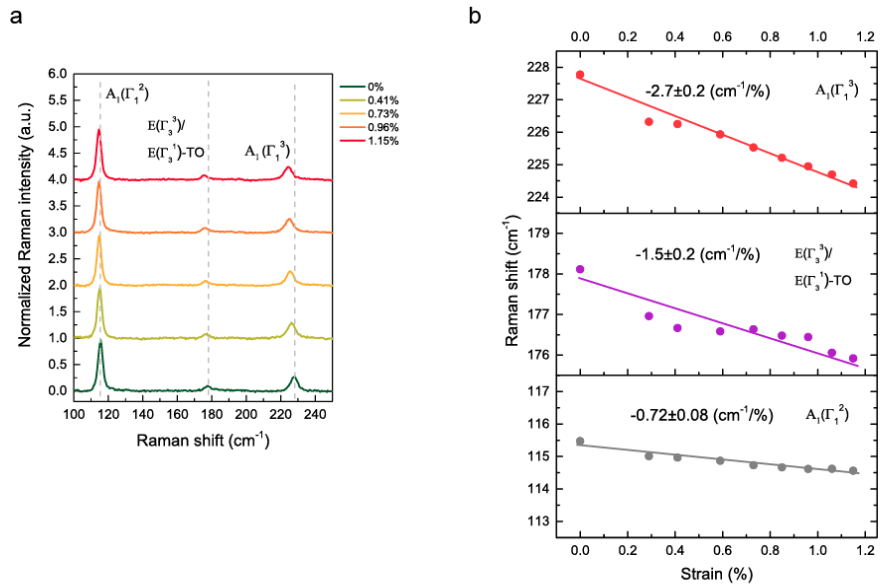

**FIG. S5.** (a) The Raman spectra of a 10-layer InSe under tensile strain with the excitation of 532nm laser. LO modes are absent in such non-resonant condition. (b) The phonon frequencies of the respective phonon modes under uniaxial tensile strain.

**S6. The PL evolution of the exciton B for strained 8-layer and 30-layer InSe**

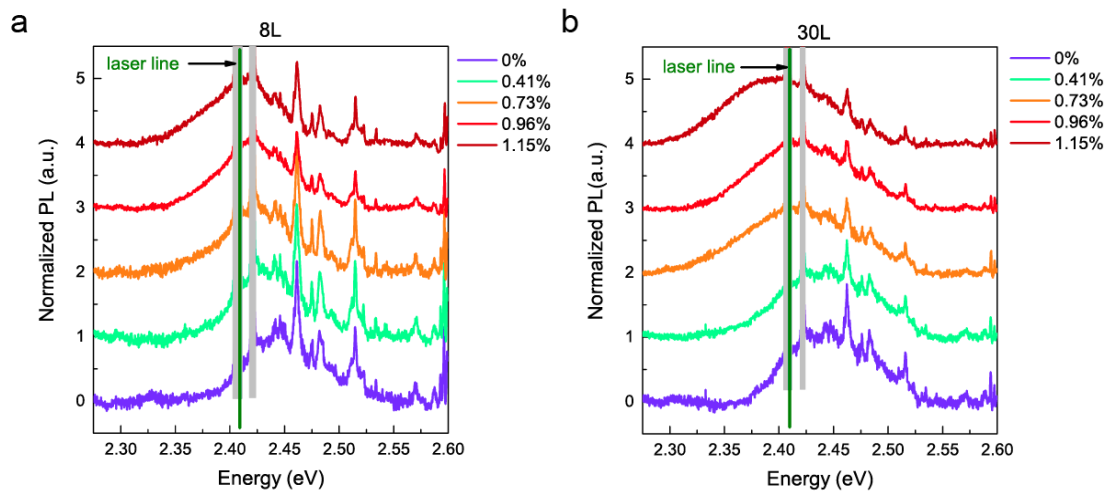

**FIG. S6.** The PL evolution of the exciton B for (a) 8-layer and (b) 30-layer InSe under uniaxial tensile strain

**S7. DFPT calculation and phonon dispersion of monolayer InSe**

To simulate the uniaxial strain effect, we expanded the crystal lattice along a-axis up to 102% with step of 0.5%. After using Vienna ab-initio simulation package (VASP) [4, 5], the atomic positions were fully optimized until all the atomic forces are smaller than 0.01 eV/Å. Then, we used VASP+phonopy strategy to calculate the phonon dispersions [6]. The force constants were calculated by density functional perturbation theory (DFPT) method. The plane-wave energy cutoff was 500 eV, and the functional was Perdew-Burke-Ernzerhof (PBE) GGA functional [7]. For energy integration, a 18×18×1 Monkhorst-Pack grid of k-points was used. Then the phonon spectra were calculated by phonopy. To account for the LO/TO splitting, the long-range electrostatic interaction was included by Non-Analytical Correction (NAC) [7]. The TO and LO splitting of $A_1$ ($\Gamma_1^1$) phonon is not seen here due to the intrinsic degeneracy in monolayer InSe, and this does not affect our discussion as our experiments find that the splitting itself allowed in few-layer InSe is quite small (<5cm$^{-1}$).

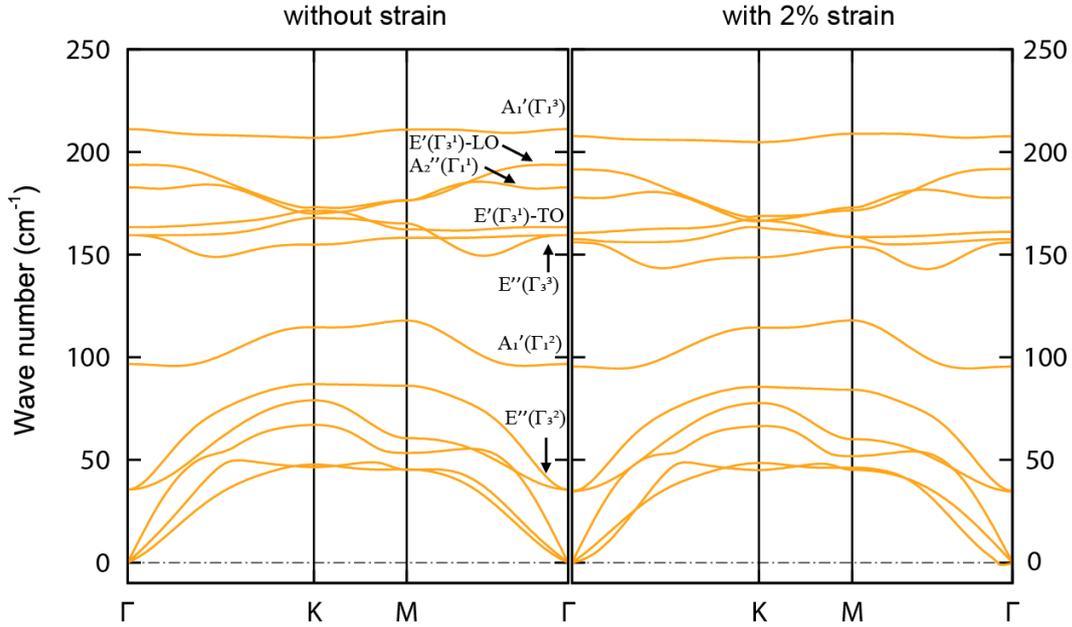

**FIG. S7.** The phonon dispersion of monolayer InSe under zero (left) and 2% (right) uniaxial tensile strain by DFPT calculations.

**S8. Calculation of the change of covalent bonds lengths under uniaxial strain**

Fig. S8 (a) and (b) show the relative change of the In-Se(a), In-Se(b) and In-In covalent bonds of monolayer InSe as uniaxial tensile strain is applied along a-axis direction. The In-Se(a) bonds (along strain direction) exhibit larger increase than the In-Se(b) bonds (perpendicular to strain direction), while the In-In bonds exhibit tiny negative change. This qualitatively explains why the shift rate of $A_1$ ($\Gamma_1^2$) phonon mode is lower than those of other modes. $A_1$ ($\Gamma_1^2$) phonon mode is the breathing mode of two In-Se sublayers, which is mainly sensitive to the out-of-plane interactions, but the out-of-plane In-In bonds exhibit a much smaller change than in-plane In-Se bonds.

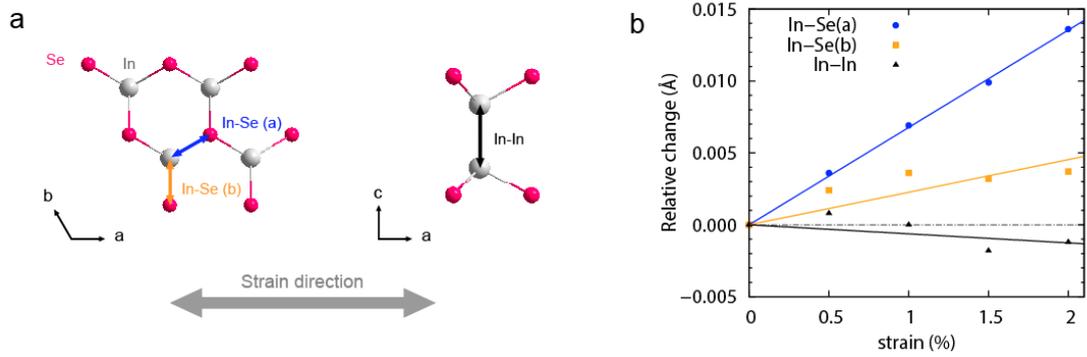

**FIG. S8.** (a) Schematic illustration of the simulated uniaxial strain direction and covalent bonds; (b) The relative change of the In-Se(a), In-Se(b) and In-In covalent bonds under strain.

**S9. The vibrational properties of strained bulk-like InSe (30-40 Layers)**

The vibrational properties of strained bulk-like (30-40 layers) InSe is generally similar to those of 10-15 layers InSe. The difference is that, as shown in Fig. S9 (a), the $A_1$ ($\Gamma_1^1$) -TO mode exhibits slight enhancement. Fig. S9 (b) demonstrates the evolution of normalized intensities of enhanced $A_1$ ($\Gamma_1^1$)-TO and LO modes. The enhancement ratio α of $A_1$ ($\Gamma_1^1$)-LO modes is 8±2, twice of that for TO mode (4±1). It should be noted that here we can only obtain the information of $A_1$ ($\Gamma_1^1$)-TO and LO phonon modes based on the spectra with two evident and separate Raman peaks. However, for some 30-40 - layer samples, the two peaks will merge together quickly and the Raman signal of $A_1$ ($\Gamma_1^1$)-TO modes will be completely masked by $A_1$ ($\Gamma_1^1$)-LO modes as tensile strain is applied. Therefore, the enhancement ratio α of $A_1$ ($\Gamma_1^1$)-LO mode (8±2) is likely underestimated. Fig. S9 (c)-(f) show the evolution of phonon frequencies of strained bulk-like InSe. The shift rates of $A_1 (\Gamma_1^2)$, $E (\Gamma_3^3) / E (\Gamma_3^1)$-TO, $A_1 (\Gamma_1^1)$-TO, $A_1 (\Gamma_1^1)$-LO and $A_1 (\Gamma_1^3)$ modes for 10-15 layers InSe are $(-0.7\pm0.1)$ cm$^{-1}$/%, $(-1.5\pm0.1)$ cm$^{-1}$/%, $(-2.3\pm0.2)$ cm$^{-1}$/%, $(-2.5\pm0.1)$ cm$^{-1}$/% and $(-2.3\pm0.1)$ cm-1/% respectively.

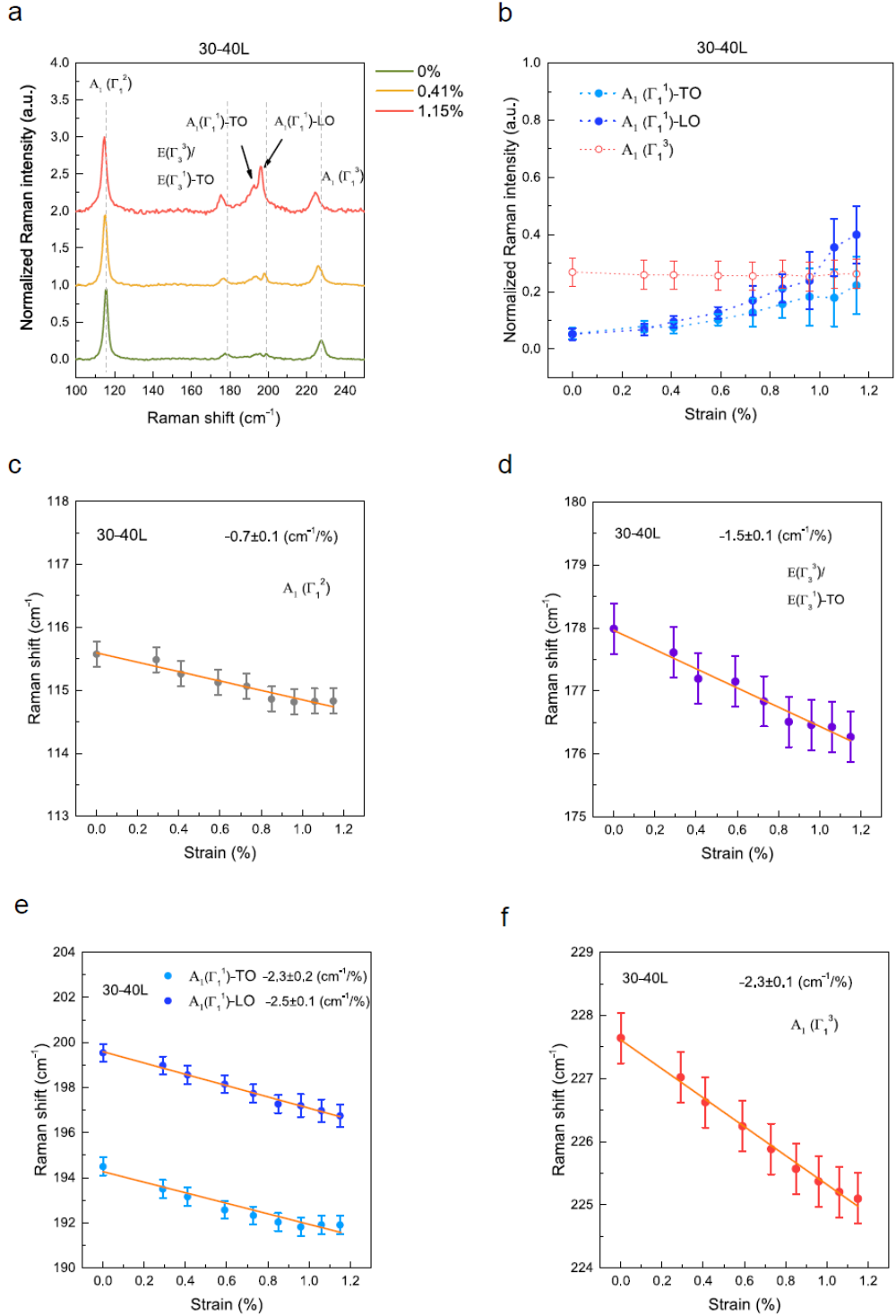

**FIG. S9.** (a) Raman spectra and (b) normalized Raman intensities of enhanced phonon

modes ($A_1$ ($\Gamma_1^1$)-TO and LO) and unenhanced $A_1$ ($\Gamma_1^2$) mode for strained 30-40 layers InSe; evolution of phonon frequencies under uniaxial tensile strain. (c)-(f) correspond to the $A_1$ ($\Gamma_1^2$), E ($\Gamma_3^3$) / E ($\Gamma_3^1$) -TO, $A_1$ ($\Gamma_1^1$)-TO, $A_1$ ($\Gamma_1^1$)-LO and $A_1$($\Gamma_1^3$) modes respectively.

## S10. Wrinkle fabrication setup and fabrication process

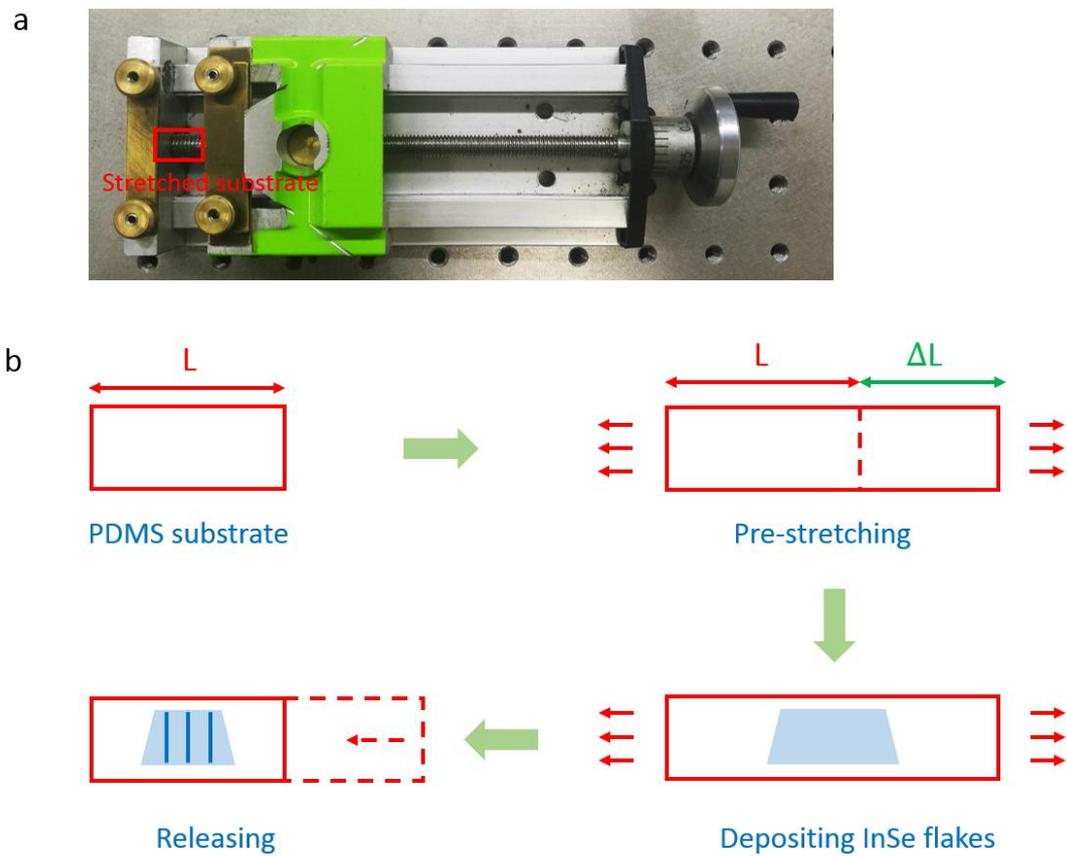

**FIG. S10.** (a) The image of our home-made wrinkle creation setup. (b) Fabrication process of wrinkled InSe.

## S11. Excitation Wavelength Dependence

To further confirm the resonance effect between the incident photon and the electronic transition, we performed wavelength-dependent Raman measurements on a 4-layer InSe. The sample was firstly exfoliated on PDMS substrate, then transferred to a silicon substrate with 300nm of silicon dioxide. An Argon laser (Stellar-Pro Select 150) with excitation wavelengths of 488nm and 514.5nm (spot size ~ 1 μm, laser power of 100 μW) was used as the excitation source. The Raman signals were recorded by a Horiba iHR550 spectrometer. The Raman peak of the silicon (at 520.7 cm$^{-1}$) substrate near the sample was used to calibrate the Raman intensity. As seen in Fig. S11 (b), the Raman intensity of $A_1$ ($\Gamma_1^1$)-LO phonon is largely enhanced with the excitation of 488nm incident photon than that of 514.5nm photon. Because the B-transition energy of 4-layer InSe is approximately 2.53eV according to our previous absorption measurements, which is resonant with the energy of the incident photon (488nm laser, photon energy 2.54eV). It is off-resonance if we use the 514.5nm laser (2.41eV).

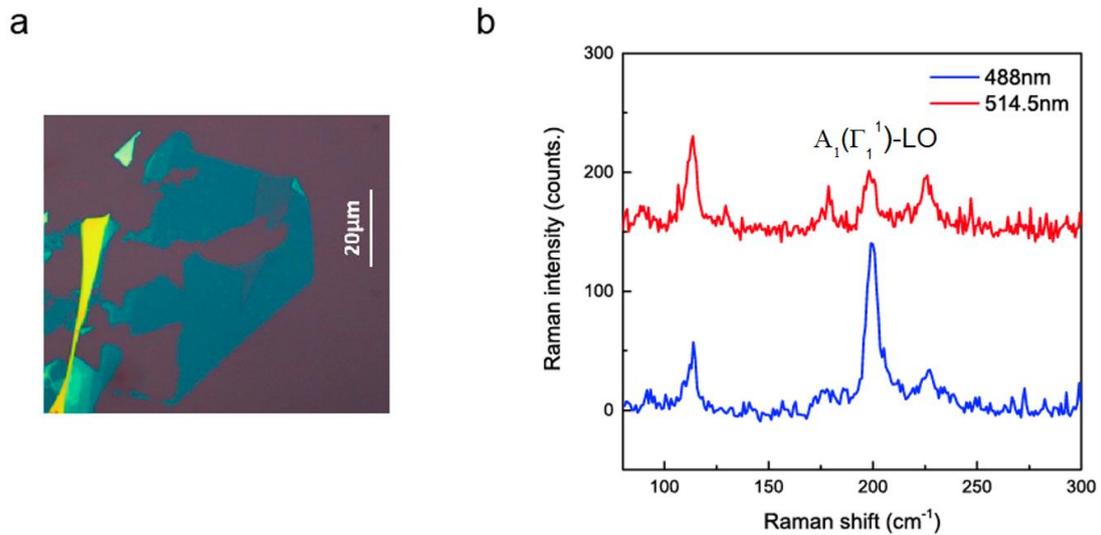

**FIG. S11.** (a) The image of a 4-layer InSe. (b) The Raman spectra with the excitation

of 488nm and 514.5nm lasers.

**S12. Comparison of the strain effect with and without clamping**

The sliding of 2D materials may occur under strain. However, few-layer InSe exhibits great stretchability owing to the much smaller Young's modulus (45 N/m for monolayer), than those of monolayer graphene (335 N/m) and MoS2 (130 N/m) [8]. Therefore, there is little sliding in moderately strained InSe, as in our experiment. To further confirm it, we performed contrast experiments to quantitatively compare the strain effect without and with clamping. We firstly applied strain and measured the Raman spectra to samples without clamping. Then, similar to the method of capping PMMA by Sujay B. Desai et al. [2], a layer of AZ5214 photoresist (with thickness of at least hundreds of nanometers) was coated on the few-layer InSe and acted as clamp, as illustrated in Fig. S12 (a). Finally, the strain was applied to the same sample and Raman spectra were taken again. Fig. S12 (b) and (c) show similar phonon frequencies of $A_1$ ($\Gamma_1^1$)-LO and $A_1$ ($\Gamma_1^3$) modes before and after clamping, in the process of applying strain. This is the case for all 6 different samples (5-15 layers) before and after clamping, and the phonon shift rates are summarized in Table S1, which all exhibit similar values and the variations are within the normal experimental uncertainty (0.1-0.3 cm$^{-1}$/%).

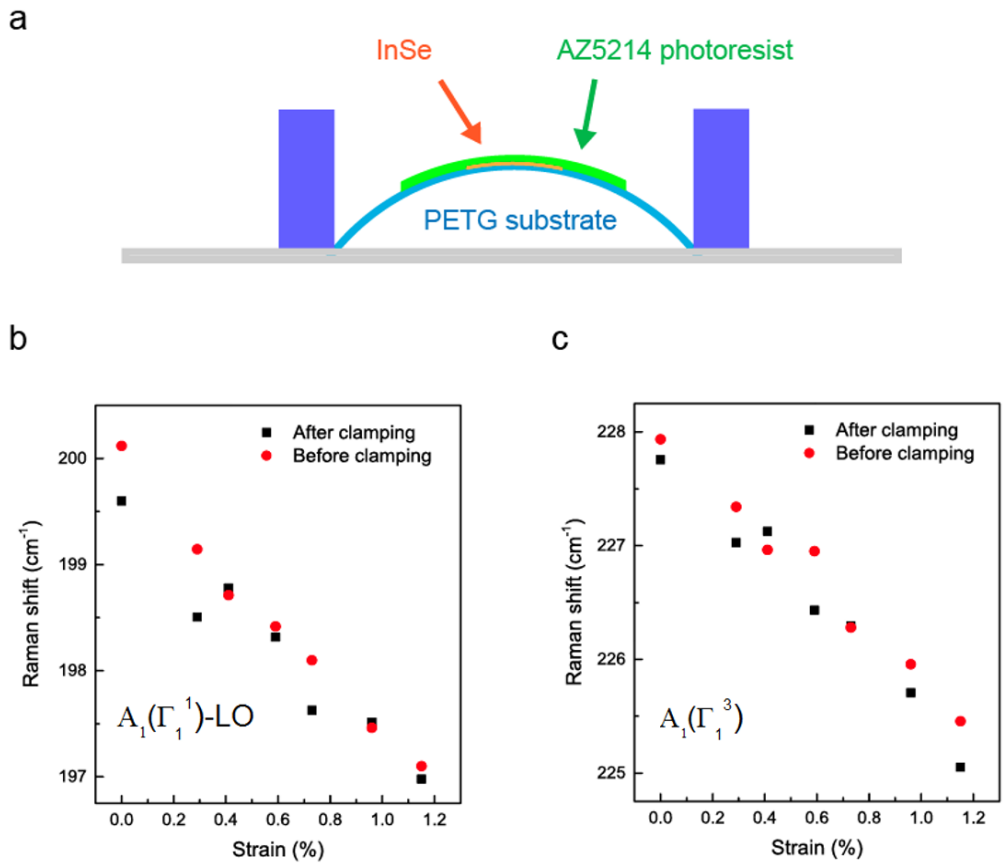

**FIG. S12** (a) Schematic illustration of clamping InSe by AZ5214 photoresist. The phonon frequencies of (b) $A_1\ (\Gamma_1^1)$-LO and (c) $A_1\ (\Gamma_1^3)$ mode for a strained InSe sample before and after clamping.

**Table S1.** The phonon shift rates before and after clamping (in cm$^{-1}$/%).

| Sample | $A_1$ ($\Gamma_1^1$)-LO mode before clamping | $A_1$ ($\Gamma_1^1$)-LO mode after clamping | $A_1(\Gamma_1^3)$ mode before clamping | $A_1(\Gamma_1^3)$ mode after clamping |
| --- | --- | --- | --- | --- |
| Sample 1 | 2.5 | 2.9 | 2.6 | 2.3 |
| Sample 2 | 2.3 | 2.3 | 2.2 | 2.5 |
| Sample 3 | 2.2 | 2.4 | 2.3 | 2.6 |
| Sample 4 | 2.0 | 2.5 | 2.2 | 2.3 |
| Sample 5 | 2.5 | 2.0 | 2.1 | 2.4 |
| Sample 6 | 2.6 | 2.2 | 2.1 | 2.3 |